\documentstyle[12pt]{article}
\begin{document}

\def\bce{\begin{center}}
\def\ece{\end{center}}
\def\beq{\begin{eqnarray}}
\def\eeq{\end{eqnarray}}
\def\ben{\begin{enumerate}}
\def\een{\end{enumerate}}
\def\ul{\underline}
\def\ni{\noindent}
\def\nn{\nonumber}
\def\bs{\bigskip}
\def\ms{\medskip}
\def\tr{\mbox{tr}}
\def\wt{\widetilde}
\def\wh{\widehat}
\def\brr{\begin{array}}
\def\err{\end{array}}
\def\dsp{\displaystyle}
\def\eg{{\it e.g.}}
\def\ie{{\it i.e.}}

\vspace*{-10mm}




\thispagestyle{empty}

\vspace*{4mm}

\begin{center}

{\large \bf  EXPLICIT ZETA FUNCTIONS FOR BOSONIC AND FERMIONIC FIELDS
ON A NONCOMMUTATIVE TOROIDAL SPACETIME}\footnote{This paper is dedicated to 
Aleix E. T., on the unique and promising occasion of his eighteenth birthday.}

\vspace{4mm}

\medskip

{\sc E. Elizalde}\footnote{E-mail: elizalde@ieec.fcr.es \
eli@ecm.ub.es \ \ http://www.ieec.fcr.es/cosmo-www/eli.html} \\
Instituto de Ciencias del Espacio (CSIC)\, \& \\  Institut d'Estudis
Espacials de Catalunya (IEEC/CSIC) \\ Edifici Nexus, Gran Capit\`{a}
2-4, 08034 Barcelona\\ and \\ Departament ECM i IFAE,
Facultat de F\'{\i}sica, \\ Universitat de Barcelona, Diagonal 647,
08028 Barcelona, Spain \\

\vspace{16mm}

{\bf Abstract}

\end{center}

Explicit formulas for the zeta functions $\zeta_\alpha (s)$ corresponding to
bosonic ($\alpha =2$) and to fermionic ($\alpha =3$) quantum fields living 
on a noncommutative, partially toroidal spacetime are derived. 
Formulas for the most  general case of the zeta function associated to a 
quadratic+linear+constant form (in {\bf Z}) are obtained. They
provide the analytical continuation of the zeta functions in question to
the whole complex $s-$plane, in terms of series of Bessel functions (of fast,
exponential convergence), thus being extended Chowla-Selberg formulas. 
As well known, this is the most convenient expression that can be
found for the analytical
continuation of a zeta function, in particular, the residua of the poles and
their finite parts are explicitly given there. An 
important novelty is the fact that
simple poles show up at $s=0$, as well as in other places (simple or double, 
depending on the
number of compactified, noncompactified, and noncommutative dimensions of the
spacetime), where they had never appeared before. This poses a challenge
to the zeta-function regularization procedure.


\vspace{2mm}


\newpage

\section{Introduction}

For its application in practice, the zeta function regularization
method relies on the existence of quite simple formulas that give
the analytical continuation of the zeta function, $\zeta (s)$, from
the region of the complex plane extending to the right of the abscissa 
of convergence,
Re $s >s_0$, to the rest of the complex plane \cite{zb1,zb2,bcvz}. 
These are not only
the reflection formula of the corresponding zeta function in each case,
but also some other, very fundamental expressions, as the Jacobi theta function
identity, Poisson's and Plana's resummation formulas, and the Chowla-Selberg
formula. However, some of these powerful expressions are often
 restricted to specific
zeta functions, and their explicit derivation is usually
quite involved. For instance, until very recently, the Chowla-Selberg (CS)
formula was only known to exist for the homogeneous, two-dimensional Epstein
zeta function. Ultimate extensions of it to more general zeta functions in
any numbeer of dimensions will be given in Sect. 2 of the present paper.

A fundamental property shared by all zeta functions is the
existence of a reflection formula. For the Riemann zeta function it is:
  $\Gamma (s/2) \zeta (s)
=\pi^{s-1/2} \Gamma (1-s/2) \zeta (1-s)$.
For a generic zeta function, $Z(s)$, we may write it as:
$
Z(\omega -s)= F(\omega, s) Z(s).
$
It allows for its analytic continuation in a very
easy way ---what is, in simple cases,
 the whole story of the zeta function regularization
 procedure. But the
analytically continued expression thus obtained
 is just another series, which has again
a slow convergence behavior, of power series type \cite{bo1} (actually the same
that the original series had, on its convergence domain).
Some years ago,  S. Chowla and A. Selberg
found a formula, for the Epstein zeta function in the two-dimensional case
 \cite{cs},
that yields {\it exponentially quick convergence everywhere}, not 
just in the reflected
domain. They were very proud of it.
In Ref. \cite{eli2a}, a first attempt was done to try to extend this 
expression  to inhomogeneous
zeta functions (very important for physical applications, see \cite{eejpa1}),
but remaining always in {\it two} dimensions,
for this was commonly believed to be an
insurmountable restriction of the original formula (see, for instance,
Ref. \cite{dic}).  More recently,
extensions to an {\it arbitrary} number of dimensions
 \cite{eecmp1,eejcam1}, both for the homogeneous (quadratic form) and
non-homogeneous (quadratic plus affine form) cases were constructed.
However, some of the new formulas (remarkably the ones corresponding
to the zero-mass case, e.g., the original CS framework!)
are {\it not explicit}, since they involve solving a rather non-trivial
recurrrence. (By the way, these explains why the CS formula had not been
extended to higher-dimensional Epstein zeta functions before.)

In  Sect. 2
we shall finish this program, by providing for the first time explicit, 
Chowla-Selberg-like
extended formulas for {\it all} possible cases involving forms of the
very general type: quadratic+linear+constant. This will complete the 
construction initiated in Refs. \cite{eecmp1,eejcam1}.

In Sect. 3 we will move to specific applications of these formulas 
in noncommutative field theory. In particular, we will obtain the explicit
analytic continuation of the zeta functions corresponding to
 scalar and vector fields defined on a quite
general, partially noncommutative toroidal manifold.
Their pole structure will be discussed in detail. The existence of
simple poles at $s=0$ comes as a  novelty in the zeta function
regularization method in this case, confirming a result obtained in Ref. 
\cite{bgz}. In other places, up to double poles will be shown to 
appear. The corresponding residua and finite parts
at the poles are immediately obtained from these expressions.

\section{Extended Chowla-Selberg formulas, associated with
arbitrary forms of quadratic+linear+constant type}

Let
$A$ a positive-definite elliptic $\Psi$DO of
positive order $m \in \mbox{\bf R}$,
 acting on
the space of smooth sections of
$E$, an $n$-dimensional vector bundle over
$M$, a  closed $n$-dimensional manifold.
The {\it zeta function} $\zeta_A$ is defined as
\beq
\zeta_A (s) = \mbox{tr}\  A^{-s} = \sum_j
 \lambda_j^{-s}, \qquad \mbox{Re}\ s>\frac{n}{m} \equiv s_0,
\eeq
where
$s_0=$ dim$\,M/$ord$\,A$ is called the {\it abscissa of
convergence} of $\zeta_A(s)$.
Under these conditions, it can be proven that
$\zeta_A(s)$ has a meromorphic
continuation to  the whole complex plane
$\mbox{\bf C}$ (regular at $s=0$),
provided that the principal symbol of $A$ (that is
$a_m(x,\xi)$) admits a {\it spectral cut}: $
L_\theta = \left\{  \lambda \in \mbox{\bf C};
 \mbox{Arg}\, \lambda =\theta,
\theta_1 < \theta < \theta_2\right\}$,   $\mbox{Spec}\, A
\cap L_\theta = \emptyset$
(Agmon-Nirenberg condition).
 The definition of $\zeta_A (s)$ depends on the
position of the cut $L_\theta$.
 The only possible singularities of $\zeta_A (s)$ are
{\it simple poles} at
$
s_k = (n-k)/m,  \  k=0,1,2,\ldots,n-1,n+1, \dots.
$
 M. Kontsevich and S. Vishik have managed to extend this
definition to the case when  $m \in \mbox{\bf C}$ (no spectral
 cut exists) \cite{kont95b}.

Consider now the following zeta function (Re $s > p/2$):
\beq
\hspace*{-3mm} \zeta_{A,\vec{c},q} (s) = {\sum_{\vec{n} \in \mbox{\bf Z}^p}}'
 \left[
\frac{1}{2}\left( \vec{n}+\vec{c}\right)^T A
\left( \vec{n}+\vec{c}\right)+ q\right]^{-s} \equiv
{ \sum_{\vec{n} \in \mbox{\bf Z}^p}}'
\left[ Q\left( \vec{n}+\vec{c}\right)+ q\right]^{-s}.  \label{zf1}
\eeq
The prime on a summation sign  means
that the point $\vec{n}=\vec{0}$ is to be excluded from the sum.
As we shall see, this is irrelevant when $q$ or some component of $\vec{c}$
is non-zero but, on the contrary, it becomes an inescapable condition
in the case when $c_1=\cdots =c_p=q=0$. Note that, alternatively, we can
view the expression inside the square brackets of the zeta function as a
sum of a quadratic, a linear, and a constant form, namely,
$Q\left( \vec{n}+\vec{c}\right)+ q = Q( \vec{n}) + L(\vec{n})+ \bar{q}$.

Our aim is to obtain a formula that gives (the analytic continuation
of) this multidimensional
zeta function in terms of an exponentially convergent series,
and which is valid in
the whole complex plane, exhibiting the singularities (poles) of
the meromorphic continuation ---with the corresponding residua---
explicitly.
 The only condition on the matrix $A$ is that
it correspond to a (non negative) quadratic form, which we call $Q$.
The vector $\vec{c}$ is arbitrary, while $q$ will be (for the
moment)  a positive constant. As we shall see, the solution to this problem
will depend very much (its explicit form) 
 on the fact that $q$ and/or $\vec{c}$ are zero or not.
According to this, we will have to distinguish different cases,
leading to unrelated final formulas,
all to be viewed as different non-trivial extensions of the CS
formula (they will be named ECS formulas, and will carry
additional tags, for the different cases).

Use of the Poisson resummation formula in Eq. (\ref{zf1}) yields
\cite{eecmp1,eejcam1}
\beq
&& \hspace*{-20mm} \zeta_{A,\vec{c},q} (s) = \frac{(2\pi )^{p/2} q^{p/2 -s}}{
\sqrt{\det A}} \, \frac{\Gamma(s-p/2)}{\Gamma (s)} +
 \frac{2^{s/2+p/4+2}\pi^s q^{-s/2 +p/4}}{\sqrt{\det A} \
\Gamma (s)}  \label{qpd1}\\ && \hspace*{2mm} \times {\sum_{\vec{m} \in
\mbox{\bf Z}^p_{1/2}}}' \cos
(2\pi
 \vec{m}\cdot \vec{c}) \left( \vec{m}^T A^{-1} \vec{m}
\right)^{s/2-p/4} \, K_{p/2-s} \left( 2\pi \sqrt{2q \,
 \vec{m}^T A^{-1} \vec{m}}\right), \nn
\eeq
where $K_\nu$ is the modified Bessel function of the second kind and
the subindex 1/2 in
$\mbox{\bf Z}^p_{1/2}$ means that in this sum, only half of the vectors
$\vec{m} \in \mbox{\bf Z}^p$ enter. That is, if we take
an $\vec{m} \in \mbox{\bf Z}^p$
we must then exclude $-\vec{m}$ (as a simple criterion one can,
for instance, select those vectors in {\bf Z}$^p \backslash \{ \vec{0}
\}$ whose first non-zero component is positive). Eq. (\ref{qpd1})
fulfills {\it all} the requirements of a CS formula.
But it is very different from the original one, constituting a non-trivial
extension to the case of a quadratic+linear+constant form, in any
number of dimensions, with the constant term being non-zero. We shall
denote this formula, Eq. (\ref{qpd1}), by the acronym ECS1.

It is notorious  how the only pole of this inhomogeneous Epstein
zeta function appears, explicitly, at $s=p/2$, where it belongs. Its
residue is given by:
\beq
\mbox{Res}_{s=p/2}  \zeta_{A,\vec{c},q} (s) =
\frac{(2\pi )^{p/2}}{\Gamma(p/2)}\, (\det A)^{-1/2}.
\eeq
\ms

\subsection{Limit  $q\rightarrow 0$}

After some work, one can obtain the limit of expression
(\ref{qpd1}) as $q\rightarrow 0$
(for simplicity we also set $\vec{c}=\vec{0}$)
 \begin{eqnarray}
&& \hspace*{-15mm}  \zeta_{A,\vec{0},0} (s) = 2^{1+s} a^{-s} \zeta (2s) +
 \sqrt{\frac{\pi}{a}}\,
\frac{\Gamma (s-1/2)}{\Gamma (s)} \,  \zeta_{\Delta_{p-1},\vec{0},0}
(s-1/2) + \frac{4 \pi^s}{a^{s/2+1/4}\, \Gamma (s)}
 \nn \\  &&  \hspace*{-12mm} \times {\sum_{\vec{n}_2 \in
\mbox{\bf Z}^{p-1}}}' \sum_{n_1=1}^\infty
 \cos \left( \frac{\pi n_1}{a}  \vec{b}^T  \vec{n}_2 \right)
n_1^{s-1/2}  \left( \vec{n}_2^T \Delta_{p-1}  \vec{n}_2
 \right)^{1/4 -s/2}
 K_{s-1/2} \left( \frac{2\pi n_1}{\sqrt{a}}
\sqrt{\vec{n}_2^T \Delta_{p-1}  \vec{n}_2 } \right). \label{cspd}
\end{eqnarray}
In Eqs. (\ref{qpd1}) and (\ref{cspd}), $A$ is a $p \times p$ symmetric matrix
$A= \left( a_{ij} \right)_{i,j=1,2, \ldots, p} =A^T$,
$A_{p-1}$  the $(p-1) \times (p-1)$ reduced submatrix
$A_{p-1}= \left( a_{ij} \right)_{i,j=2, \ldots, p}$,
$a$ the first component, $a=a_{11}$, $\vec{b}$ the $p-1$ vector
$\vec{b} =(a_{21}, \ldots, a_{p1})^T = (a_{12}, \ldots, a_{1p})^T$, and
 $\Delta_{p-1}$ is the following $(p-1) \times (p-1)$  matrix:
 $\Delta_{p-1} =  A_{p-1}- \frac{1}{4a} \vec{b} \otimes \vec{b}$.
 More precisely, what one actually obtains by taking the limit is {\it 
the reflected formula}, as one would get after using the Epstein 
zeta function reflection
$
\Gamma (s) Z(s;A) = \pi^{2s-p/2} (\det A)^{-1/2}
 \Gamma (p/2-s) Z(p/2-s;A^{-1}),
$
being $Z(s;A)$ the Epstein zeta function \cite{eps1}. Finally. it can be
written as (\ref{cspd}). (It is a rather non-trivial exercise to perform
this calculation.) Note that
Eq. (\ref{cspd}) has {\it all} the properties demanded from a CS formula,
but it is actually {\it not explicit}. It is in fact a recurrence, rather
lengthy to solve as it stands. In fact, it can be viewed as the 
{\it straightforward
extension} of the original CS formula to higher dimensions. It was the
top result of previous work on this subject, for the case 
$q=c_1=\cdots =c_p=0$ \cite{eecmp1,eejcam1}.

Using a different strategy, this recurrence will be now
solved explicitly, in a much more simple way. Indeed, let us proceed in
a complementary way, namely, by doing  the inversion provided by the
Poisson resummation formula (or the Jacobi identity), with respect to
$p-1$ of the indices (say, $j=2,3, \ldots, p$). This leaves us with
three sums, corresponding to positive, zero, and negative values of the
remaining
index ($n_1$, in this case). The zero value of $n_1$ (in correspondence with
the rest of the $n_i$'s not being all zero) classifies the number of
different situations (according to the values of the  $c_i$'s an $q$ being
all zero or not) into just two cases. (As is immediate, from start all
$c_i$'s can be taken to be between 0 and 1:
$0\leq c_i < 1, \, i=1,2,\ldots, p$.)\, (i) The first case is, thus,
when at least  one of the
$c_i$'s or $q\geq 0$ is not zero. Since the case $q\neq 0$ has been solved
already, we will mean by this case now that, say $c_1\neq 0$. (ii)
The second case is when all $q=c_1=\cdots =c_p=0$.
\ms

\subsection{Case with $q=0$ but $c_1\neq 0$}
\ms

\subsubsection{General (non-diagonal) subcase}

By doing the inversion provided by the
Poisson resummation formula (or the Jacobi identity), with respect to
$p-1$ of the indices (here, $j=2,3, \ldots, p$), we readily obtain:
 \begin{eqnarray}
&& \hspace*{-6mm}  \zeta_{A_p,\vec{c},0} (s) = \frac{2^{s}}{\Gamma (s)} \,
 \left(\det{A_{p-1}} \right)^{-1/2} \left\{  \pi^{(p-1)/2} \left( a_{11}-
\vec{a}_{p-1}^T A_{p-1}^{-1}\vec{a}_{p-1} \right)^{(p-1)/2-s}
\Gamma \left( s-(p-1)/2 \right) \right. \nn
\\  && \hspace*{4mm} \times \,
\left[ \zeta_H (2s-p+1,c_1) + \zeta_H (2s-p+1,1-c_1)\right] + 4
\pi^s \left( a_{11}-\vec{a}_{p-1}^T
A_{p-1}^{-1}\vec{a}_{p-1} \right)^{(p-1)/4-s/2} \nn
\\  && \hspace*{-7mm}\times  \sum_{n_1 \in \mbox{\bf Z}}
{\sum_{\vec{m} \in \mbox{\bf Z}^{p-1}_{1/2}}}' \cos \left[ 2\pi
 \vec{m}^T \left( \vec{c}_{p-1}+ A_{p-1}^{-1}\vec{a}_{p-1}(n_1+c_1 )
\right)\right] \, |n_1+c_1|^{(p-1)/2-s} \left( \vec{m}^TA_{p-1}^{-1}
\vec{m}\right)^{s/2-(p-1)/4} \nn
\\  &&  \hspace*{2mm}\left. \times \, K_{(p-1)/2-s}\left(2\pi |n_1+c_1| \sqrt{
\left( a_{11}-\vec{a}_{p-1}^T A_{p-1}^{-1}\vec{a}_{p-1} \right)
\vec{m}^TA_{p-1}^{-1}\vec{m}}\right)\right\} - \left( \frac{1}{2}
\vec{c}^TA\vec{c}
\right)^{-s}.
 \label{ecsc1}
\end{eqnarray}
Here, and in what follows,
 $A_{p-1}$ is (as before) the submatrix of $A_p$ composed of the last
$p-1$ rows and columns. Moreover, $a_{11}$ is  the first diagonal
component of $A_p$,
while $\vec{a}_{p-1} =(a_{12}, \ldots, a_{1p})^T =(a_{21}, \ldots, a_{p1})^T$,
and $\vec{m}=(n_2,\ldots, n_p)^T$. Note that this is an {\it explicit formula},
that the only pole at $s=p/2$ appears also explicitly, and that the second term
of the rhs is a series of exponentially fast convergence. It has,
therefore (as Eq. (\ref{qpd1})), all the properties required to qualify as
a CS formula. We shall name this expression ECS2.

\ms

\subsubsection{Diagonal subcase}

In this very common, particular case the preceding expression reduces to
the more simple form:
 \begin{eqnarray}
  \zeta_{A_p,\vec{c},0} (s)& =& \frac{2^{s}}{\Gamma (s)} \,
 \left(\det{A_{p-1}} \right)^{-1/2} \left\{  \pi^{(p-1)/2}  a_{1}^{(p-1)/2-s}
\Gamma \left( s-(p-1)/2 \right) \,\right. \nn
\\  && \hspace*{40mm} \times \,
\left[ \zeta_H (2s-p+1,c_1) + \zeta_H (2s-p+1,1-c_1)\right]  \nn
\\  &&  \hspace*{-14mm} + 4 \pi^s a_{1}^{(p-1)/4-s/2} \sum_{n_1 \in
\mbox{\bf Z}}
{\sum_{\vec{m} \in \mbox{\bf Z}^{p-1}_{1/2}}}' \cos \left( 2\pi
 \vec{m}^T  \vec{c}_{p-1}\right) \, |n_1+c_1|^{(p-1)/2-s}  \left(
\vec{m}^TA_{p-1}^{-1}\vec{m}\right)^{s/2-(p-1)/4}\nn
\\  && \times \, \left. K_{(p-1)/2-s}\left(2\pi |n_1+c_1| \sqrt{
 a_{1}\vec{m}^TA_{p-1}^{-1}\vec{m}}\right)\right\} - \left( \frac{1}{2}
 \vec{c}^TA\vec{c} \right)^{-s}.   
\label{ecsc2}
\end{eqnarray}
We shall call this formula ECS2d.
\ms

\subsection{Case with $c_1=\cdots =c_p=q=0$}

\ms

\subsubsection{General (non-diagonal) subcase}

As remarked in \cite{eecmp1,eejcam1}, we had not been able to obtain here
yet a closed formula, but just a (rather non-trivial) recurrence, Eq. 
(\ref{cspd}), relating the $p-$dimensional case with
the $(p-1)-$dimensional one. After a second look, we have now
realized that we can actually
still proceed as if we had in fact  $c_1=1 \neq 0$, both for
positive and for negative values of $n_1$. A sum, though, remains with
$n_1=0$ ---and the rest of the $n_i$'s not all being zero--- what
yields, once more, the same zeta function of the begining, but corresponding 
to $p-1$ indices. All in all:
 \begin{eqnarray}
  \zeta_{A_p,\vec{0},0} (s) &=& \zeta_{A_{p-1},\vec{0},0} (s)+
\frac{2^{1+s}}{\Gamma (s)} \, \left(\det{A_{p-1}} \right)^{-1/2} \left\{
  \pi^{(p-1)/2}\left( a_{11}-
\vec{a}_{p-1}^T A_{p-1}^{-1}\vec{a}_{p-1} \right)^{(p-1)/2-s}
 \right. \nn
\\  && \hspace*{65mm}
 \times \,\Gamma \left( s-(p-1)/2 \right) \, \zeta_R (2s-p+1)
\label{ecsc3}
\\  &&  \hspace*{-15mm} + 4 \pi^s \left( a_{11}-\vec{a}_{p-1}^T
A_{p-1}^{-1}\vec{a}_{p-1} \right)^{(p-1)/4-s/2} \sum_{n=1}^\infty
{\sum_{\vec{m} \in \mbox{\bf Z}^{p-1}_{1/2}}}' \cos \left( 2\pi
 \vec{m}^T A_{p-1}^{-1}\vec{a}_{p-1}n \right) \, n^{(p-1)/2-s} \nn
\\  &&  \hspace*{-6mm} \left. \times \, \left( \vec{m}^TA_{p-1}^{-1}
\vec{m}\right)^{s/2-(p-1)/4} K_{(p-1)/2-s}\left[2\pi n \sqrt{
\left( a_{11}-\vec{a}_{p-1}^T A_{p-1}^{-1}\vec{a}_{p-1} \right)
\vec{m}^TA_{p-1}^{-1}\vec{m}}\right]\right\}.
 \nn
\end{eqnarray}
This is also a recurrent expression, an alternative to (\ref{cspd}), obtained
with the help of a different strategy.

Remarkably enough, it is easy to resolve this recurrence explicitly,
and indeed to obtain a {\it closed formula} for this case (we shall write the
dimensions of the submatrices of $A$ as subindices). The result being
 \begin{eqnarray}
  \zeta_{A_p} (s) &\equiv & \zeta_{A_p,\vec{0},0} (s) =
\frac{2^{1+s}}{\Gamma (s)} \sum_{j=1}^{p} \left(\det{A_{p-j}}
\right)^{-1/2} \left\{
  \pi^{(p-j)/2}\left( a_{jj}-
\vec{a}_{p-j}^T A_{p-j}^{-1}\vec{a}_{p-j} \right)^{(p-j)/2-s}
 \right. \nn
\\  && \hspace*{55mm}
 \times \,\Gamma \left( s-(p-j)/2 \right) \, \zeta_R (2s-p+j)
\label{ecsc33}
\\  &&  \hspace*{-18mm} + 4 \pi^s \left( a_{jj}-\vec{a}_{p-j}^T
A_{p-j}^{-1}\vec{a}_{p-j} \right)^{(p-j)/4-s/2} \sum_{n=1}^\infty
{\sum_{\vec{m}_{p-j} \in \mbox{\bf Z}^{p-j}_{1/2}}}' \cos \left( 2\pi
 \vec{m}_{p-j}^T A_{p-j}^{-1}\vec{a}_{p-j}n \right) \, n^{(p-j)/2-s} \nn
\\  &&  \hspace*{-17mm} \left. \times \, \left( \vec{m}_{p-j}^TA_{p-j}^{-1}
\vec{m}_{p-j}\right)^{s/2-(p-j)/4} K_{(p-j)/2-s}\left[2\pi n \sqrt{
\left( a_{jj}-\vec{a}_{p-j}^T A_{p-j}^{-1}\vec{a}_{p-j} \right)
\vec{m}_{p-j}^TA_{p-j}^{-1}\vec{m}_{p-j}}\right]\right\}.
 \nn
\end{eqnarray}
With a similar notation as above, here $A_{p-j}$ is the submatrix of $A_p$
composed of the last $p-j$ rows and columns. Moreover, $a_{jj}$ is
 the j-th diagonal component of $A_p$,
while $\vec{a}_{p-j} =(a_{jj+1}, \ldots, a_{jp})^T =(a_{j+1j}, \ldots,
a_{pj})^T$, and $\vec{m}_{p-j}=(n_{j+1},\ldots, n_p)^T$.
Again, this is an extension of the Chowla-Selberg formula to the case
in question. It exhibits all the same good properties. Physically, it
corresponds to the homogeneous, massless case. It is
 to be viewed, in fact,
as {\it the} genuine multidimensional extension of the
Chowla-Selberg formula. We shall call it ECS3.

\ms

\subsubsection{Diagonal subcase}

Let us particularize once more to the diagonal case,  with
$\vec{c}=\vec{0}$, which is quite important in practice and gives
rise to more simple expressions. For the recurrence, we have
 \begin{eqnarray}
&& \hspace*{-10mm}  \zeta_{A_p} (s) = \zeta_{A_{p-1}} (s)+
\frac{2^{1+s}}{\Gamma (s)} \, \left(\det{A_{p-1}} \right)^{-1/2} \left[
  \pi^{(p-1)/2} a_{1}^{(p-1)/2-s}
\Gamma \left( s-(p-1)/2 \right) \, \zeta_R (2s-p+1)
 \right. \label{ecsc4}
\\  &&  \hspace*{-10mm}\left. + 4 \pi^s  a_{1}^{(p-1)/4-s/2}
\sum_{n=1}^\infty
{\sum_{\vec{m} \in \mbox{\bf Z}^{p-1}}}'  n^{(p-1)/2-s}
\left( \vec{m}^TA_{p-1}^{-1}\vec{m}\right)^{
s/2-(p-1)/4} K_{(p-1)/2-s}\left(2\pi n \sqrt{
 a_{1}\vec{m}^TA_{p-1}^{-1}\vec{m}}\right)\right]. \nn
\end{eqnarray}
As above,
we can solve  this finite recurrence and obtain the following simple and
explicit formula for this case:
 \begin{eqnarray}
&& \hspace*{-16mm}  \zeta_{A_p} (s)  = 
\frac{2^{1+s}}{\Gamma (s)} \,
\sum_{j=0}^{p-1} \left(\det{A_j} \right)^{-1/2} \left[ \pi^{j/2}
a_{p-j}^{j/2-s} \Gamma \left( s-j/2 \right) \, \zeta_R(2s-j)  \right.
\nn \\  &&  \hspace*{-9mm}\left. + 4 \pi^s a_{p-j}^{j/4-s/2}
\sum_{n=1}^\infty
{\sum_{\vec{m}_j \in \mbox{\bf Z}^j}}' n^{j/2-s}
\left(\vec{m}_j^t A_j^{-1}\vec{m}_j\right)^{s/2-j/4}
K_{j/2-s}\left(2\pi n \sqrt{ a_{p-j} \vec{m}_j^t A_j^{-1}\vec{m}_j}\right)
\right],\label{ecs1}
\end{eqnarray}
with $A_p = \mbox{diag} (a_1, \ldots, a_p)$, $A_j = \mbox{diag} (a_{p-j+1},
\ldots, a_p)$, $\vec{m}_j =(n_{p-j+1}, \ldots, n_p)^T$, and $\zeta_R$  the
Riemann zeta function. Note again the fact that this and
Eq.  (\ref{ecsc33}) are {\it explicit}
expression for the multidimensional, generalized Chowla-Selberg formula and,
in this way, they go beyond any result obtained previously.
We name this formula ECS3d.

It is immediate to see that the term for $j=0$ in the sum yields the last term,
$\zeta_{A_1} (s)$, of the recurrence, that is:
 \begin{eqnarray}
\zeta_{A_1} (s) = {\sum_{n_p=-\infty}^{+\infty}}' \left( \frac{a_p}{2}
n_p^2 \right)^{-s} = 2^{1+s}a_p^{-s} \zeta_R(2s).
\end{eqnarray}
It exhibits a pole, at $s=1/2$ which is spurious ---it is actually {\it not}
a pole of the whole function (since it cancels, in fact, with another
one coming from the next term, with further cancellations of this kind
going on, each term with the next).
Concerning the pole structure of the resulting zeta function, as given
by Eq. (\ref{ecs1}), it is not difficult to see that {\it only} the pole
at $s=p/2$ is actually there (as it should). It is in the last term, $j=p-1$,
of the sum, and it has the correct  residue, namely
 \begin{eqnarray}
\left. \mbox{Res } \zeta_{A_p} (s)\right|_{s=p/2}  = \frac{(2\pi)^{p/2}}{
\Gamma(p/2)} \,  \left(\det{A_p} \right)^{-1/2}. \label{res11}
\end{eqnarray}
The rest of the seem-to-be poles at $s=(p-j)/2$ are not such: they
compensate among themselves, one term of the sum with the next,
adding pairwise to zero.

Summing up, this formula, Eq. (\ref{ecs1}),  provides a convenient
analytic continuation of the zeta function to the whole complex
plane, with its only simple pole showing up explicitly. Aside from
this, the finite part of the first sum in the expression is quite
easy to obtain, and the remainder ---an awfully looking multiple
series--- is in fact an extremely-fast convergent sum, yielding a
 small contribution, as happens in the
CS formula. In fact, since it corresponds to the
case $q=0$, this expression should be viewed as {\it the}
extension of the original  Chowla-Selberg formula ---for the zeta
function associated with an homogeneous quadratic form in two
dimensions--- to an arbitrary number, $p$, of dimensions. The rest
of the formulas above provide also extensions of the original CS expression.

The general case of a quadratic+linear+constant form has been here
thus completed. As we clearly see, the main formulas
corresponding to the three different subcases, namely
ECS1 Eq. (\ref{qpd1}), ECS2 Eq. (\ref{ecsc1}), and ECS3 Eq. (\ref{ecsc33}), 
are in fact quite distinct and one cannot directly go from one to another 
by adjusting some parameters.

For the sake of completeness, we must mention the following.  Notice that 
all cases considered here correspond to having
a non-identically-zero quadratic form $Q$. For $Q$ identically zero, that is, 
the linear+constant (or affine) case, the formulas for the analytic 
continuation are again quite different from the ones 
above. The corresponding zeta function is called Barnes' zeta function.
This case has been throughly studied in Ref. \cite{eejcam1}.
\bs

\section{Spectral zeta function for both scalar and vector fields on a
spacetime with a noncommutative toroidal part}

We shall now consider the physical example of a quantum system consisting 
of scalars and vector fields on a
$D-$dimensional noncommutative manifold, $M$, of the form
$\mbox{\bf R}^{1,d}\bigotimes \mbox{\bf T}_\theta^p$ (thus $D=d+p+1$).
$\mbox{\bf T}_\theta^p$ is a $p-$dimensional noncommutative torus,
its coordinates satisfying the usual relation: $[x_j,x_k]=i\theta
\sigma_{jk}$. Here $\sigma_{jk}$ is a real nonsingular, antisymmetric 
matrix of $\pm 1$ entries, and $\theta$ is 
the noncommutative parameter. 

This physical system has attracted much interest
recently, in connection with $M-$theory and with string theory
\cite{connes,douglas,seiberg,cheung,chu,schomerus,ardalan}, and
also because of the fact that those are perfectly consistent
theories by themselves, which could lead to brand new physical situations. 
It has been  shown, in particular, that
noncommutative gauge theories describe the low energy excitations
of open strings on $D-$branes in a background Neveu-Schwarz
two-form field \cite{connes,douglas,seiberg}. 

This interesting
system provides us with a quite non-trivial case where the
formulas derived above are indeed useful. For one, the zeta
functions corresponding to bosonic and fermionic operators in this
system are of a different kind, never considered before. And, moreover,
they can be most conveniently written in terms of the zeta functions
in the previous section. What is also nice is the fact that a unified
treatment (with just {\it one} zeta function) can be given for both cases,
the nature of the field appearing there as a simple parameter, together with
those corresponding to the numbers of compactified, noncompactified, 
and noncommutative dimensions of the spacetime. 

\subsection{Poles of the zeta function}

The spectral zeta function for the corresponding
(pseudo)differential operator can be written in the form \cite{bgz}
\begin{eqnarray}
\hspace*{-4mm} \zeta_{\alpha} (s) = \frac{V}{(4\pi )^{(d+1)/2}}
\frac{\Gamma (s-(d+1)/2)}{\Gamma (s)} {\sum_{\vec{n} \in \mbox{\bf
Z}^p}}' Q (\vec{n})^{(d+1)/2-s} \left[ 1+ \Lambda
\theta^{2-2\alpha}Q (\vec{n})^{-\alpha}\right]^{(d+1)/2-s},
\label{za1}
\end{eqnarray}
where $V=$ Vol\,($\mbox{\bf R}^{d+1})$, the volume of the
non-compact part, and $Q (\vec{n}) = \sum_{j=1}^p a_j n_j^2$, a
diagonal quadratic form, being the compactification radii
$R_j=a_j^{-1/2}$. Moreover, the value of the parameter $\alpha =2$ for 
scalar fields and $\alpha =3$ for vectors, distinguishes between the
two different fields. In the particular case when we set all
the compactification radii equal to $R$, we obtain:
\begin{eqnarray}
\hspace*{-4mm} \zeta_{\alpha} (s) = \frac{V}{(4\pi )^{(d+1)/2}}
\frac{\Gamma (s-(d+1)/2)}{\Gamma (s) R^{d+1-2s}} {\sum_{\vec{n}
\in \mbox{\bf Z}^p}}' I (\vec{n})^{(d+1)/2-s} \left[ 1+ \Lambda
\theta^{2-2\alpha} R^{2\alpha} I
(\vec{n})^{-\alpha}\right]^{(d+1)/2-s}, \label{za11}
\end{eqnarray}
being now the quadratic form: $I (\vec{n}) = \sum_{j=1}^p n_j^2$.

After some calculations, this zeta function can be written in
terms of the Epstein zeta function of the previous section, with the
result:
\begin{eqnarray}
\zeta_{\alpha} (s) = \frac{V}{(4\pi )^{(d+1)/2}} \sum_{l=0}^\infty
\frac{\Gamma (s+l-(d+1)/2)}{l! \, \Gamma (s)}\, (-\Lambda
\theta^{2-2\alpha})^l \ \zeta_{Q,\vec{0},0}(s+\alpha l-(d+1)/2),
\label{za2}
\end{eqnarray}
which reduces, in the particular case of equal radii, to
\begin{eqnarray}
\zeta_{\alpha} (s) = \frac{V}{(4\pi )^{(d+1)/2}R^{d+1-2s}}
\sum_{l=0}^\infty \frac{\Gamma (s+l-(d+1)/2)}{l! \, \Gamma (s)}\,
(-\Lambda \theta^{2-2\alpha})^l \ \zeta_E(s+\alpha l-(d+1)/2),
\label{za21}
\end{eqnarray}
where we use here the notation $\zeta_E (s)\equiv
\zeta_{I,\vec{0},0} (s)$, e.g., the Epstein zeta function for the
standard quadratic form.

The pole structure of the resulting zeta function deserves a careful
analysis. It differs, in fact, very much from all cases that were
known in the literature till now. This is not difficult to understand, 
from the fact that the pole of the Epstein zeta function at
$s=p/2-\alpha k+(d+1)/2=D/2-\alpha k$, when combined with the poles of the
gamma functions, yields a very rich pattern of singularities for
$\zeta_\alpha (s)$, on taking into account the different possible 
values of the 
parameters involved. The pole structure is straightforwardly found 
from the explicit expressions for the zeta functions in Sect. 2.

Having already given the formula (\ref{za2}) above ---that contains
everything needed to perform such calculation of pole position, residua
and finite part--- for its importance for the calculation of 
the determinant and
the one-loop effective action from the zeta function,  we will here start by
specifying what happens at $s=0$.  Remarkably enough, a pole
appears in many cases (depending on the values of the different parameters). 
This will also serve as an illustration of what
one has to expect for other values of $s$. The general case will be left 
for the following subsection.

It is convenient to classify the different possible subcases according to
the values of $d$ and $D=d+p+1$. We obtain, at $s=0$, the pole
structure given in Table 1.

\begin{table}[bht]

\begin{eqnarray}
\mbox{For } \, d=2k \qquad \left\{ \begin{array}{ccc} \mbox{if }
\, D\neq \dot{\overline{2\alpha}} & \Longrightarrow & \zeta_\alpha
(0) =0, \\   \mbox{if } \, D = \dot{\overline{2\alpha}} &
\Longrightarrow & \zeta_\alpha (0) =\mbox{finite}. \end{array}
\right. \nonumber
\end{eqnarray}
\begin{eqnarray}
\mbox{For } \, d=2k-1 \quad \left\{ \begin{array}{ccc} \mbox{if }
\, D\neq \dot{\overline{2\alpha}} \  \left\{ \begin{array}{ll}
\mbox{finite}, & \mbox{for } \, l\leq k \\ 0, & \mbox{for } \, l
> k
\end{array} \right\} & \Longrightarrow & \zeta_\alpha (0) =\mbox{finite},
\\ \mbox{} \\ \mbox{if } \, D = 2\alpha l \  \left\{ \begin{array}{ll}
\mbox{pole}, & \mbox{for } \, l\leq k \\  \mbox{finite}, &
\mbox{for } \, l > k
\end{array} \right\}  & \Longrightarrow &
\zeta_\alpha (0) =\mbox{pole}. \end{array} \right. \nonumber
\end{eqnarray}

\caption{{\protect\small Pole structure of the zeta function
$\zeta_\alpha (s)$, at $s=0$, according to the different possible
values of $d$ and $D$ ($\dot{\overline{2\alpha}}$ means {\it multiple} of 
$2\alpha$.)}}

\end{table}

Here $l$ is the summation index in Eq. (\ref{za2}). The appearance
of a pole of the zeta function $\zeta_\alpha (s)$, for both values
of $\alpha$, at $s=0$ is, let us repeat, an absolute novelty, bound to have
important physical consequences for  the regularization
process. It is necessary to observe, that this fact is {\it not} in
contradiction with the well known theorems on the pole structure
of a (elliptic) differential operator \cite{gil1}. The situation that appears
in the noncommutative case is completely different. (i) To begin with,
we do not have any longer a standard differential operator, but a
strictly {\it pseudodifferential} one, from the beginning. (ii) Moreover, the 
new spectrum is {\it not}
perturbatively connected (for $\theta \to 0$) with the corresponding
one for the commutative case. \ms

\subsection{Explicit analytic continuation of $\zeta_\alpha (s),
\, \alpha =2,3$, in the complex $s-$plane}

 Substituting the corresponding formula, from the
preceding section, for the Epstein zeta functions in Eq.
(\ref{za2}), we obtain the following explicit analytic
continuation of $\zeta_\alpha (s)$ ($\alpha =2,3$),
for bosonic and fermionic fields, to the {\it whole} complex
$s-$plane:
\begin{eqnarray}
\zeta_{\alpha} (s) &=& \frac{2^{s-d} \, V}{(2\pi )^{(d+1)/2}\Gamma
(s)} \sum_{l=0}^\infty \frac{\Gamma (s+l-(d+1)/2)}{l! \, \Gamma
(s+\alpha l -(d+1)/2)}\, (-2^\alpha \Lambda \theta^{2-2\alpha})^l
\sum_{j=0}^{p-1} \left(\det{A_{j}} \right)^{-1/2} \nn
\\ && \times \,  \left[ \pi^{j/2} a_{p-j}^{-s-\alpha l
+(d+j+1)/2} \Gamma (s+\alpha l-(d+j+1)/2) \zeta_R(2s+2\alpha
l-d-j-1)  \right. \nn \\ && \quad \ \, + \, 4 \pi^{s+\alpha
l-(d+1)/2} a_{p-j}^{-(s+\alpha l)/2-(d+j+1)/4} \sum_{n=1}^\infty
{\sum_{\vec{m}_j \in \mbox{\bf Z}^j}}' n^{(d+j+1)/2-s-\alpha l}
 \label{za4} \\ && \left. \qquad  \times \, \left(\vec{m}_j^t
A_j^{-1}\vec{m}_j\right)^{(s+\alpha l)/2-(d+j+1)/4}
K_{(d+j+1)/2-s-\alpha l}\left(2\pi n \sqrt{ a_{p-j} \vec{m}_j^t
A_j^{-1}\vec{m}_j}\right) \right]. \nn
\end{eqnarray}

As discussed in the previous subsection in detail, the
non-spurious poles of this zeta function are to be found in the
terms corresponding to $j=p-1$. With the knowledge we have gained
from the analytical continuation of the Epstein zeta functions in
Sect. 2, the final analysis can be here completed at once. 
Note that the situation here corresponds to the case of
Subsect. 2.3.2, namely, the diagonal case with $c_1=\cdots
=c_p=q=0$.

To be remarked again is that, what we have in the end, by using
our method, is an exponentially fast convergent series of Bessel
functions together with a first, finite part, where a 
pole (simple or double, as we shall see) may show up, for specific 
values of the dimensions of the
different parts of the manifold, depending also on the nature
(scalar vs. vectorial) of the fields (the value of  $\alpha$, see
Table 1 and  Eq. (\ref{za4})).

To summarize the discussion at the end of Sect. 2, the pole
structure of Eq. (\ref{za4}) is in fact best seen from Eq.
(\ref{za2}) (for $s=0$ it has been analyzed in Table 1 already). For a fixed
value of the summation index $l$, the contribution to the only
pole of the zeta function $\zeta_E (s+\alpha l-(d+1)/2)$, at
$s=D/2-\alpha l$, comes from the last term of the $j-$sum only,
namely from $j=p-1$. It is easy to check that it yields the
corresponding residuum (\ref{res11}). This corresponds to the
second sum in Eq. (\ref{za4}). Combined now with the poles of the
gamma functions, and taking into account the first series in $l$,
this yields the following expression for the residua of the zeta
function $\zeta_\alpha (s)$ at the poles $s=D/2-\alpha l$,
$l=0,1,2, \ldots$
 \begin{eqnarray}
 \left. \mbox{Res } \zeta_\alpha  (s)\right|_{s=D/2-\alpha l}
  &=& \frac{2^{p/2-d} \pi^{(p-d-1)/2} V}{
\Gamma(p/2)} \,  \left(\det{A_p} \right)^{-1/2} \,  \frac{(-
\Lambda \theta^{2-2\alpha})^l}{l!} \, \frac{\Gamma (p/2
+(1-\alpha)l)}{\Gamma (D/2 -\alpha l)}, \nonumber \\ &&  l=0,1,2,
\ldots \label{res21}
\end{eqnarray}
Actually, depending on $D$ and $p$ being even or odd, completely
different situations arise, for different values of $l$: from
the disappearance of the pole, giving rise to a finite
contribution, to the appearance of a simple or a double pole. We shall
distinguish four different situations and, to simplify the
notation, we will denote by $U$ the whole factor in the
expression (\ref{res21}) for the residuum, that multiplies the last
fraction of two gamma functions (in short, Res $\zeta_\alpha =U \,
\Gamma_1/\Gamma_2$).
\begin{enumerate}
\item For $D-2\alpha l= -2h$, \ $h=0,-1,-2, \ldots$
\begin{enumerate}
\item for $p/2+(1-\alpha) l \neq 0,-1,-2, \ldots \qquad \Longrightarrow
$ \qquad finite, \quad \\ Res $\zeta_\alpha = -h! \ U \ \Gamma (p/2
+(1-\alpha)l)$;
\item for $p=2(\alpha -1) l -2k$, $k = 0,-1,-2, \ldots \qquad \Longrightarrow
$ \qquad pole, \quad \\ Res $\zeta_\alpha = (h!/k!) \ U $.
\end{enumerate}
\item For $D-2\alpha l \neq -2h$, $h= 0,-1,-2, \ldots$
\begin{enumerate}
\item for $p/2+(1-\alpha) l \neq 0,-1,-2, \ldots \qquad \Longrightarrow
$ \qquad pole, \quad \\ Res $\zeta_\alpha =  U \ \Gamma (p/2
+(1-\alpha)l)/\Gamma (D/2 +\alpha l) $;
\item for $p=2(\alpha -1) l -2k$, $k = 0,-1,-2, \ldots \qquad \Longrightarrow
$ \qquad double pole, \quad \\ Res $\zeta_\alpha = (-1/k!) \ U \,
/\Gamma (D/2 +\alpha l) $.
\end{enumerate}
\end{enumerate}
Note that we here just quote the {\it generic} situation that occurs for
$l$ {\it large} enough in each case. For instance, if $p=2$ a double
pole appears for $l=1,2,\ldots$. For $p=4$, a double pole appears
for $l=1,2,\ldots$, if $\alpha=3$, but only for $l=2,3,\ldots$, if
$\alpha=2$. For $p=6$, a double pole appears for $l=2,3,\ldots$,
if $\alpha=3$, but only for $l=3,4,\ldots$, if $\alpha=2$, and so
on. The case with both $D$ and $p$ even (what implies $d$ odd) is
the most involved one. For $p=2$ and $D=4$, for instance, there is
a transition from a pole for $l=0$ corresponding to the zeta
function factor, to a pole for $l=1$ and higher, corresponding to the
gamma function in the numerator (the compensation of the pole of
the zeta function factor with the one coming from the gamma function in
the denominator prevents the formation of a double pole). In any
case, the explicit analytic continuation of $\zeta_\alpha (s)$
given by Eq. (\ref{za4}) contains {\it all} the information one needs for
calculating the poles and corresponding  residua in a straightforward way.

The pole structure can be summarized as in Table 2.

\begin{table}[htb]

\begin{center}

\begin{tabular}{|c||c|c|}
\hline \hline p \ $\backslash$ \ D &  even &   odd  \\
 \hline \hline
 odd & (1a) \ \, {\it pole} / finite ($l\geq l_1$) & (2a) \ \, 
{\it pole} / pole  \\ \hline
 even & (1b) \ \, {\it double pole} / pole ($l\geq l_1,l_2$) & (2b) \
 \,
{\it pole} / double pole ($l\geq l_2$)
 \\ \hline
 \hline \end{tabular}

\caption{{\protect\small General pole structure of the zeta
function $\zeta_\alpha (s)$, according to the different possible
values of $D$ and $p$ being odd or even. In italics, the type of behavior
corresponding to lower values of $l$ is quoted, while the behavior shown
in roman characters corresponds to larger values of $l$. }}

\end{center}

\end{table}

An application of these formulas to the calculation of the
one-loop partition function corresponding to quantum fields at finite 
temperature, on a noncommutative flat spacetime, will be given elsewhere
\cite{bez1}.

\ms

\noindent{\bf Acknowledgments}

Thanks are given to Andrei Bytsenko and Sergio Zerbini for stimulating
discussions on this subject, and to the members of the Physics Department, 
University of Trento, for the warm hospitality.
This investigation has been supported by DGI/SGPI (Spain), project
BFM2000-0810, by CIRIT (Generalitat de Catalunya),
contract 1999SGR-00257, and by the program INFN (Italy)--DGICYT (Spain).

\vspace{5mm}


\end{document}